\documentclass[preprint,showpacs,superscriptaddress,aps]{revtex4-1}
\usepackage{amssymb}
\usepackage{amsmath,amssymb,graphicx}
\usepackage{graphicx}
\usepackage{dcolumn}
\usepackage{bm}
\usepackage{color}
\def\be{\begin{equation}}
\def\ee{\end{equation}}
\def\bea{\begin{eqnarray}}
\def\eea{\end{eqnarray}}

\def\lsim{\raise0.3ex\hbox{$\;<$\kern-0.75em\raise-1.1ex\hbox{$\sim\;$}}}
\def\gsim{\raise0.3ex\hbox{$\;>$\kern-0.75em\raise-1.1ex\hbox{$\sim\;$}}}

\def\slash#1{\setbox0=\hbox{$#1$}#1\hskip-\wd0\dimen0=5pt\advance
\dimen0 by-\ht0\advance\dimen0 by\dp0\lower0.5\dimen0\hbox
to\wd0{\hss\sl/\/\hss}} \hoffset=0.0cm \voffset=0.0cm

\usepackage{pstricks}
\begin{document}

\title{Charged Higgs contribution to  $\bar{B}_s \rightarrow \phi \pi^0 $ and $\bar{B}_s \rightarrow \phi \rho^0 $ }

\author{Gaber Faisel}
\email{gfaisel@hep1.phys.ntu.edu.tw}
\affiliation{{\fontsize{10}{10}\selectfont{Department of Physics,
National Taiwan University, Taipei, Tawian 10617.}}}
\affiliation{{\fontsize{10}{10}\selectfont{Department of Physics
and Center for Mathematics and Theoretical Physics, National
Central University, Chung-Li, Tawian 32054.}}}
\affiliation{{\fontsize{10}{10}\selectfont{Egyptian Center for
Theoretical Physics, Modern University for Information and
Technology, Cairo, Egypt.}}}

\begin{center}

\begin{abstract}
We study the decay modes $\bar{B}_s\to \phi \pi^0$ and
$\bar{B}_s\to \phi \rho^0$ within the frameworks of two-Higgs
doublet models type-II and typ-III. We adopt in our study Soft
Collinear Effective Theory as a framework for the calculation of
the amplitudes. We  derive the contributions of the charged Higgs
mediation to the weak effective Hamiltonian governing the decay
processes in both models. Moreover we  analyze the effect of the
charged Higgs mediation on the Wilson coefficients of the
electrowek penguins and on the branching ratios of $\bar{B}_s\to
\phi \pi^0$ and $\bar{B}_s\to \phi \rho^0$ decays. We show that
wthin two-Higgs doublet models type-II and type-III the Wilson
coefficients corresponding to the electroweak penguins can be
enhanced due to the contributions from the charged Higgs mediation
leading into enhancement in the branching ratios of $ \bar{B}_s\to
\phi \pi^0$ and $\bar{B}_s\to \phi \rho^0 $ decays.  We find that,
within two-Higgs doublet models type-II, the enhancement in the
branching ratio of $\bar{B}_s\to \phi \pi^0$ can not exceed $18\%$
with respect to the SM predictions. For the branching ratio of
$\bar{B}_s\to \phi \rho^0$, we find that the charged Higgs
contribution in this case is small where the branching ratio of
$\bar{B}_s\to \phi \rho^0$ can be enhanced or reduced by about 4\%
with respect to the SM predictions.  For the case of the two-Higgs
doublet models type-III we show that the branching ratio of
$\bar{B}_s\to \phi \pi^0$ can be enhanced by about a factor $2$ of
its value within two-Higgs doublet models type-II. However no
sizable enhancement with respect to the SM predictions can be
obtained for both $\bar{B}_s\to \phi \pi^0$ and $\bar{B}_s\to \phi
\rho^0$ decays.

\end{abstract}
\end{center}

\maketitle
\section{ Introduction}

Within Standard Model (SM) flavour-changing neutral current (FCNC)
decays are generated at the one loop level. As a result they are
highly suppressed  and can serve as a sensitive probe of possible
New Physics(NP) beyond SM. Of particular interest are the purely
isospin-violating decays $\bar B_s\to\phi\rho^0$ and $\bar
B_s\to\phi\pi^0$ that are dominated by electroweak penguins
\cite{Fleischer:1994rs}. They have been studied within SM in
different  frameworks such as  QCD factorization as in
Refs.\cite{Beneke:2003zv,Hofer:2010ee}, in PQCD  as in
Ref.\cite{Ali:2007ff} and using Soft Collinear Effective Theory
(SCET) as in Refs.\cite{Wang:2008rk,Faisel:2011kq}. In
Ref.\cite{Hofer:2010ee} the study has been extended to include NP
models namely, a modified $Z^0$ penguin, a model with an
additional $U(1)'$~ gauge symmetry and the MSSM using QCDF. Their
results showed that the additional $Z\,'$ boson of the $U(1)'$~
gauge symmetry with couplings to leptons  switched off can enhance
the electroweak penguin amplitude sizably  leading to an
enhancement in their branching ratios by up to an order of
magnitude. This finding makes these decay modes are very
interesting for LHCb and future $B$ factories searches
\cite{Hofer:2010ee}.   Motivated by this possibility we  extend
the study  to the  two Higgs doublet models (2HDMs).

  In 2HDMs, the Higgs sector of the SM  can be extended to include extra
$SU(2)_L$ scalar doublet. Accordingly, the simplest picture of the
SM Higgs coupling to the quarks and leptons can be modified by the
presence of the extra Higgs doublet. This results in several
classes of 2HDMs such as 2HDMs type-I, type-II, type-III, type-X
and type-Y \cite{Haber:1978jt,Abbott:1979dt,barger,
KY,Crivellin:2012ye,Crivellin:2013wna}.  For 2HDMS type-I and
type-II an investigation of the effect of the charged Higgs
contributions to the electrweak penguins  has been done in
Ref.\cite{Grossman:1999av} where the interest was to explore their
significance to $B\to K\pi$ decay modes. Their conclusion is that
the significant contributions to the electrweak penguins are
favored for small charged Higgs mass and $\cot\beta=1$. However
taking into account $B\to X_s\gamma$ constraints rule out this
possibility.

  In the present work we derive the new contributions to the
electrweak penguins that are proportional to $m_b \tan^2\beta/m_t$
which were neglected in Ref.\cite{Grossman:1999av}. These new
contributions become dominant when $\tan\beta$ becomes large as we
will show in the following. Moreover the charged Higgs mediation
at tree-level can lead to a set of new operators that can not be
generated in the SM. We derive their contributions to the
effective Hamiltonian governs the process under consideration and
calculate their corresponding Wilson coefficients.  Having all
these new contributions we will give the predictions for the
branching ratios of $\bar{B}_s\to \phi \pi^0$ and $\bar{B}_s\to
\phi \rho^0$ within 2HDMs type-II which has not been calculated in
Ref.\cite{Grossman:1999av}. In addition we extend our study to
include  2HDMs type-III which has generic Yukawa structure that
can allow for sizable effects in FCNC processes as shown in Ref.
\cite{Crivellin:2013wna} and can also enhance CP violation in
charm sector \cite{Delepine:2012xw}.

 In this work we adopt SCET as a framework for the calculation
of the
amplitudes\cite{Bauer:2000ew,Bauer:2000yr,Chay:2003zp,Chay:2003ju}.
SCET provides a systematic and  rigorous way to deals with the
processes in which energetic quarks and gluons have different
momenta modes such as hard, soft and collinear modes. The power
counting in SCET  reduces the complexity of the calculations.  In
addition, the factorization formula given by SCET is perturbative
to all powers in $\alpha_s$ expansion.

 This paper is organized as follows. In Sec.~\ref{sec:formalism},
we briefly review the decay amplitude for $B \to M_1M_2$ within
SCET framework. Accordingly, we give a brief review of the SM
contribution to the branching ratios of $ \bar{B}_s\to \phi \pi^0$
and $ \bar{B}_s\to \phi \rho^0 $ decays within SCET framework.
Then we derive the Wilson coefficients in the case of Two
Higgs-doublets models type II and type III and analysis  their
contributions to the branching ratios of $ \bar{B}_s\to \phi
\pi^0$ and  $ \bar{B}_s\to \phi \rho^0 $ in section ~\ref{SMHeff}.
Finally, we give our conclusion in Sec.~\ref{sec:conclusion}.

 \section{ $ B\to {M_1M_2}$ in $ SCET $ }\label{sec:formalism}

 At leading order in $\alpha_s$ expansion, the amplitude of $ B\to
{M_1M_2}$ where $M_1$ and $M_2$ are light mesons  can be written
as

 \begin{eqnarray}
 {\cal A}^{LO}_{B\rightarrow M_1M_2}&=&\frac{G_F
m_B^2}{\sqrt{2}}\Big(f_{M_1}\bigg[\int^1_0
 dudzT_{M_1J}(u,z)\zeta^{B M_2}_J(z)\phi_{M_1}(u)\nonumber\\
  &+& \zeta^{B M_2}\int^1_0
 du T_{M_1\zeta}(u)\phi_{M_1}(u)\bigg]+(M_1 \leftrightarrow
 M_2)\Big).\label{amp1}
 \end{eqnarray}

 The hard kernels $T_{(M_1,M_2)\zeta}$ and $T_{(M_1,M_2) J}$
can be expressed  in terms of the  Wilson coefficients depending
on the final states mesons $M_1$ and  $M_2$. We refer to Refs.
\cite{Bauer:2004tj,Williamson:2006hb} for  explicit expressions of
$T_{(M_1,M_2)\zeta}$ and $T_{(M_1,M_2) J}$ for different $M_1$ and
$M_2$ final states mesons. The hadronic parameters $\zeta^{BM}$
and $\zeta_J^{BM}$ that appear in Eq.(\ref{amp1}) are related to
the form factors for  $B\to M$ transitions through the combination
$\zeta^{BM}+\zeta_J^{BM}$\cite{Bauer:2005kd}. The power counting
implies that $\zeta^{BM}\sim \zeta^{BM}_J\sim
(\Lambda/m_b)^{3/2}$\cite{Bauer:2005kd}. Generally, we expect to
have large number of $\zeta^{BM}$ and $\zeta_J^{BM}$ for the 87
$B\to PP$ and $B\to VP$ decay channels.  However, using symmetries
like SU(2) and SU(3) can reduce the number of these parameters
\cite{Bauer:2005kd,Williamson:2006hb,Wang:2008rk}. On the other
hand a model independent analysis requires to determine them from
the experimental  data as done for few decay modes of B mesons in
Refs.\cite{Bauer:2005kd,Jain:2007dy}.  For a large number of $B$
and $B_s$ decays, the $\chi^2$ fit method, using the experimental
data of the branching fractions and CP asymmetries  of the  non
leptonic $B$ and $B_s$ decays, have been used in
Refs.\cite{Williamson:2006hb,Wang:2008rk} to determine
$\zeta^{BM}$ and $\zeta_J^{BM}$. We refer to
refs.\cite{Bauer:2005kd,Jain:2007dy} for details about the fit
method to determine $\zeta^{BM}$ and $\zeta_J^{BM}$.

 In our analysis, we follow ref.\cite{Williamson:2006hb} and assume a  $20\%~$ error in
both $\zeta^{B(M_1,M_2)}$ and $\zeta_J^{B (M_1,M_2)}$  due to  the
SU(3) symmetry breaking. In addition, we use  the values of
$\zeta^{B(M_1,M_2)}$ and $\zeta_J^{B (M_1,M_2)}$ given in
ref.\cite{Wang:2008rk}  corresponding to the two solutions
obtained from the $\chi^2$ fit.  For the light cone distribution
amplitudes we use the same input values given in
ref.\cite{Jain:2007dy}. Following our work in
Ref.\cite{Faisel:2011kq}, the amplitudes of $\bar{B}_s\to \phi
\pi^0$ and $\bar{B}_s\to \phi \rho^0$ decays corresponding to
solution 1 of the SCET parameters are given as \bea
 {\cal A}(\bar{B}^0_s\to \phi \pi^0)\times
10^6&\simeq& (-3.6 C_{10} + 1.4 \tilde{C}_{10} + 8.3 C_7 - 8.3
\tilde{C}_{7} + 1.9 C_8 - 1.9 \tilde{C}_{8} - 8.3 C_9 + 6.6
\tilde{C}_{9})\lambda^s_{t}\nonumber\\ &+& (2.4 C_1 - 0.9
\tilde{C}_{1} + 5.6 C_2 -
4.4\tilde{C}_{2})\lambda^s_{u}\nonumber\\{\cal A}(\bar{B}_s\to
\phi\rho^0 )\times 10^6&\simeq& (-8.3 C_{10} -4.3  \tilde{C}_{10}
-11.9 C_7 +11.9 \tilde{C}_{7} + 0.4  C_8 - 0.4\tilde{C}_{8} -11.9
C_9 + 0.05 \tilde{C}_{9})\lambda^s_{t}\nonumber\\ &+& (5.5 C_1+
2.9 \tilde{C}_{1} + 7.9 C_2 - 0.03 \tilde{C}_{2})\lambda^s_{u}
\label{rho1}\label{pi1} \eea

while for solution 2 of the SCET parameters we have
\cite{Faisel:2011kq}

 \bea
 {\cal A}(\bar{B}^0_s\to \phi \pi^0)\times
10^6&\simeq& (-5.1 C_{10} - 0.3 \tilde{C}_{10} + 9.3 C_7 - 9.3
\tilde{C}_{7} + 1.1 C_8 - 1.1 \tilde{C}_{8} - 9.3 C_9 + 5.2
\tilde{C}_{9})\lambda^s_{t}\nonumber\\ &+& (3.4 C_1 + 0.2
\tilde{C}_{1} + 6.2 C_2 -
3.4\tilde{C}_{2})\lambda^s_{u}\nonumber\\{\cal A}(\bar{B}_s\to
\phi\rho^0 )\times 10^6&\simeq& (-7.4 C_{10} +0.33  \tilde{C}_{10}
-14.9 C_7 +14.9 \tilde{C}_{7} - 2.5  C_8 + 2.5 \tilde{C}_{8} -
14.9  C_9 + 8.3 \tilde{C}_{9})\lambda^s_{t}\nonumber\\ &+& (4.9
C_1- 0.22  \tilde{C}_{1} + 9.9 C_2 -5.5
\tilde{C}_{2})\lambda^s_{u} \label{pi2}\eea

here $C_i $ and $\tilde{C}_i $ are the Wilson coefficients that
can be  expressed as \be C_i = C_i^{SM} +
C_i^{H^{\pm}},~~~~~~~~~~~~~~~~~~~~~~~~~~~~~\tilde{C}_i =
\tilde{C}_i^{H^{\pm}} \ee

$\tilde{C}_i $ are the Wilson coefficients corresponding to
four-quark operators in the weak effective Hamiltonian that can be
obtained by flipping the chirality from left to right and so in
the SM  $\tilde{C}_i^{SM}=0$.  It should be noted that the
expressions of the  amplitude of  $\bar{B}_s\to \phi\rho^0$
considered above is only for the decay of $\bar{B}_s$ to  two
longitudinally polarized $\phi$ and $\rho^0$ mesons. At leading
order  in the $1/m_b$ expansion  expansions, one can match the
weak effective Hamiltonian at the scale $\mu \sim m_b$ for $\Delta
S =1$ two body $B$ decays to a $SCET_I$ Hamiltonian. The $SCET_I$
Hamiltonian can be expressed in terms of two set of operators
namely the leading order operators $Q^{(0)}_{if}$ and the relevant
subleading operators $Q^{(1)}_{if}$ in the $\sqrt{\lambda}$
expansion \cite{Williamson:2006hb}. Here $f$ refer to $d$ and $s$
quarks and $i=1,2,..$. These are the only relevant operators as
higher order operators will be suppressed due to the smallness of
the scaling parameter $\lambda$ that is defined as $\lambda=
\Lambda_{QCD}/m_b$.  The  decay of $\bar{B}_s$ to  two
transversely polarized mesons,  $\bar{B}_s\to V_\perp V_\perp$, do
not receive contributions from $Q^{(0)}_{if}$ and $Q^{(1)}_{if}$
operators and thus the amplitude given in Eq.(\ref {amp1}) is for
$PP$, $PV$ and for two longitudinally polarized vector mesons,
$B\to V_\parallel V_\parallel$, \cite{Williamson:2006hb}. Here $P$
and $V$ stands for pseduscalar and vector mesons respectively.

  In  Refs.\cite{Beneke:2005we,Lu:2006nza} it was pointed out that
$\bar{B}\to V_\perp V_\perp$  decays can be enhanced by  the
presence of an enhanced ${\mathcal O} (m_b) $ electromagnetic
operator. This operator can lead to a contribution that are
$m_b/\Lambda$ enhanced compared to the amplitudes for $B\to
V_\parallel V_\parallel$, but which are, on the other hand, also
$\alpha^{\rm em}$ suppressed due to the exchanged photon
\cite{Williamson:2006hb}. Thus, numerically, the contribution from
the electromagnetic operator can be expected to be smaller than
the ${\mathcal O} (m_b^0)$ terms in Eq.(\ref{amp1})
\cite{Williamson:2006hb}. Hence at leading order the only
contributions to $B\to V_\perp V_\perp$ can arise from
nonperturbative charming penguins $A_{cc}$ \cite{Bauer:2004tj},
which does not contribute to $\bar{B}_s\to \phi\rho^0$ decay,
while the other terms are either $1/m_b$ or $\alpha_0^{\rm
em}m_b/\Lambda$ suppressed \cite{Williamson:2006hb}.

  The predictions for the branching ratios of $\bar{B}^0_s\to \phi
\pi^0$ and $\bar{B}_s\to \phi \rho^0$ within SM are presented in
Table \ref{branch}. As can be seen from Table \ref{branch}, the
SCET predictions for the branching ratios are smaller than PQCD
and QCDF predictions. This can be explained as the predicted form
factors in SCET are smaller than those used in PQCD and
QCDF\cite{Wang:2008rk}.

\begin{table}
\begin{center}
\begin{tabular}{|c|c|c|c|c|c|c|}
  \hline
  Decay channel & QCD factorization & PQCD & SCET solution $1$ & SCET solution $2$\\
  \hline
  $\bar{B}_s\to \phi \pi^0$  & $16_{-3}^{+11}$ & $ 16 _{-5-2-0}^{+6+2+0}$&
  $ 7_{-1-2}^{+1+2} $&$ 9 _{-1-4}^{+1+3}$ \\
  $\bar{B}_s\to \phi \rho^0$  & $ 44 _{-7}^{+27}$ & $23_{-7-1-1}^{+9+3+0}$
  & $ 20.2^{+1+9}_{-1-12}$ & $ 34.0^{+1.5 + 15}_{-1.5-22}$  \\
  \hline
\end{tabular}
 \end{center}
\caption{ Branching ratios in units $10^{-8}$ of $ \bar{B}_s\to
\phi \pi^0$ and $\bar{B}_s\to \phi \rho^0$  decays. The last two
columns  give the predictions corresponding to the amplitudes in
Eqs.(\ref{pi1},\ref{pi2})\cite{Faisel:2011kq}.
 On the SCET predictions the errors are
due to  the CKM matrix elements  and  SU(3) breaking effects
respectively. For a comparison with previous studies in the
literature, we list the results evaluated in QCDF
\cite{Hofer:2010ee}, PQCD \cite{Ali:2007ff}.}\label{branch}
\end{table}

 As can be seen from Table \ref{branch}, the
branching ratios of $\bar{B}^0_s\to \phi \rho^0$ are larger  than
the branching ratios of $\bar{B}^0_s\to \phi \pi^0$. Both
$\bar{B}^0_s\to \phi \rho^0$ and $\bar{B}^0_s\to \phi \pi^0$
decays  are generated via the $\bar{B}_s \to \phi$ transition.
Thus they have the same non perturbative form factors
$\zeta^{B\phi}$ and $\zeta_J^{B\phi}$. However, using a
non-polynomial model for the light cone distribution amplitude
$\phi_{\rho}(u)$ in the case of  $\bar{B}^0_s\to \phi \rho^0$
decay can lead to a slightly different result from using the
polynomial model for the light cone distribution amplitude
$\phi_{\pi}(u)$ in the case of $\bar{B}^0_s\to \phi \pi^0$ decay
as pointed out in ref.\cite{Jain:2007dy}. Another reason for this
difference is that the Wilson coefficients $C_7 $ and $C_8 $ enter
the hard kernels, $ T_{1\zeta}(u)$ and $ T_{1 J}(u,z) $ of
$\bar{B}^0_s\to \phi \rho^0$ with opposite signs to the case in
$\bar{B}^0_s\to \phi \pi^0$\cite{Faisel:2011kq}.

\section{Models with Charged Higgs bosons}\label{SMHeff}

Charged Higgs can exist as one of the new Higgs particles in any
possible extension of the Higgs sector of the SM such as two Higgs
doublet models.  In the literature,  the 2HDM of type II has been
investigated in many processes due to its simple Yukawa sector
which respects flavor conservation by requiring that one Higgs
doublet couple  to down type-quarks and charged leptons while the
other one couples to up-type quarks only such as the Higgs
potential of the MSSM and so on. One way to  achieve this is by
imposing a symmetry on the Lagrangian such as $Z_2$ symmetry.
Clearly, in the 2HDM of type II there are no FCNC at tree level
can be induced by exchanging neutral Higgs particles and flavor
violation can be induced only by the CKM matrix elements entering
the charged Higgs vertex.

  In the two Higgs doublet models type-III both Higgs can couple to
up and down type quarks and upon taking some limits we restore
back two Higgs doublet model type-II as we will show in the
following. Thus the Yukawa sector of this model will allow for
FCNC at tree level not only by the charged Higgs mediation but
also with the exchanging of neutral Higgs particles. One can avoid
the unwanted FCNC at tree level by imposing strong constraints on
the new couplings from several observables in some processes as we
show in the following. However some new couplings can still escape
these constraints and thus can lead to interesting results as
explaining the $B\to D^* \tau \nu$ anomaly which can not be
explained in 2HDMs type-II \cite{Crivellin:2012ye}. In addition
these new coupling can be in general complex and thus  can lead to
 new sources of weak CP violating phases which can enhancedirect
 CP asymmetries comparing to the SM.

 The Yukawa Lagrangian of  the 2HDMs type-III
can be written as \cite{Crivellin:2010er,Crivellin:2012ye} :
\begin{eqnarray}
\mathcal{L}^{eff}_Y &=& \bar{Q}^a_{f\,L} \left[
  Y^{d}_{fi} \epsilon_{ab}H^{b\star}_d\,-\,\epsilon^{d}_{fi} H^{a}_u \right]d_{i\,R}\\
&-&\bar{Q}^a_{f\,L} \left[ Y^{u}_{fi}
 \epsilon_{ab} H^{b\star}_u \,+\, \epsilon^{ u}_{fi} H^{a}_d
  \right]u_{i\,R}\,+\,\rm{h.c}. \,,\nonumber\label{Yuka}
\end{eqnarray}
where $\epsilon_{ab}$ is the totally antisymmetric tensor, and
$\epsilon^q_{ij}$ parameterizes the non-holomorphic corrections
which couple up (down) quarks to the down (up) type Higgs doublet.
After electroweak symmetry breaking the two Higgs doublets $H_u$
and $H_d$ result in  the physical Higgs mass eigenstates $A^0$
(CP-odd Higgs),  $H^0$ (heavy CP-even Higgs), $h^0$ (light CP-even
Higgs) and $H^{\pm}$. In our study we follow
Refs.\cite{Crivellin:2010er,Crivellin:2012ye} and assume a
MSSM-like Higgs potential and thus the charged Higgs mass is given
by
\begin{eqnarray}
m^2_{H^{\pm}} = m^2_{A^0}+m^2_W \label{mH}
\end{eqnarray}

 where the $W$ boson mass, $m_W$, is related to the  the vacuum expectation values of
the neutral component of the  Higgs doublets, $v_u$ and $v_d$, via

\begin{eqnarray}
m^2_W = \frac{1}{2}g^2(v^2_u+v^2_d)= \frac{1}{2}g^2 v^2
\end{eqnarray}
and the mass $m_{A^0}$ is treated as a free parameter. It should
be noted that in the limit $v << m_{A^0} $ all heavy Higgs masses
($m_{H^0}$ , $m_{A^0}$ and $m_{H^{\pm}}$) are approximately equal
\cite{Crivellin:2013wna}.

   The effective Lagrangian $\mathcal{L}^{eff}_Y$  gives
rise to   the following charged Higss-quarks interaction
Lagrangian:
\begin{equation}
\mathcal{L}^{eff}_{H^\pm} = \bar{u}_f {\Gamma_{u_f d_i
}^{H^\pm\,LR\,\rm{eff} } }P_R d_i
+ \bar{u}_f {\Gamma_{u_f d_i }^{H^\pm\,RL\,\rm{eff} } }P_L d_i\, ,\\
 \label{Higgs-vertex}
\end{equation}
with \cite{Crivellin:2012ye} \bea {\Gamma_{u_f d_i
}^{H^\pm\,LR\,\rm{eff} } } &=& \sum\limits_{j = 1}^3 {\sin\beta\,
V_{fj} \left( \frac{m_{d_i }}{v_d} \delta_{ji}-
  \epsilon^{ d}_{ji}\tan\beta \right), }
\nonumber\\
{\Gamma_{u_f d_i }^{H^ \pm\,RL\,\rm{eff} } } &=& \sum\limits_{j =
1}^3 {\cos\beta\,  \left( \frac{m_{u_f }}{v_u} \delta_{jf}-
  \epsilon^{ u\star}_{jf}\tan\beta \right)V_{ji}}
 \label{Higgsv}
\eea Here  $V$ is the CKM matrix and  tan $\beta = v_u/v_d$. Using
the Feynman-rule given in Eq.(\ref{Higgs-vertex}) we can derive
the contributions of the charged Higgs mediation to the weak
effective Hamiltonian governs the $b\to s$ transition. The weak
effective Hamiltonian in this case is generated from diagrams
similar to the case of the SM with the replacing of the charged
$W$ bosons with the charged Higgs bosons. Thus the weak effective
Hamiltonian  is the same as in the SM with only exception is that
the presence of a new set of operators obtained from the SM ones
by changing the chirality from left to right. For the left
chirality operators we derived the corresponding  Wilson
coefficients due to the charged Higgs mediation and we find that
they are given as:
\begin{eqnarray}
C_{1,2}^{(H^{\pm})}&=&0,\nonumber\\
C_3^{(H^{\pm})}&=& - \frac { \sqrt{2}\,\alpha_s\cos^2\beta }{24
\pi G_F m^2_{H^{\pm}}} \left( \frac{m_t }{v_u} -\epsilon ^{
u\,\star}_{33}\tan\beta \right)} { \left(\frac{m_t}{v_u}
-\epsilon^u_{33}\tan\beta \right)
  I_1(x),\nonumber\\
C_4^{(H^{\pm})}&=&  \frac { \sqrt{2}\,\alpha_s\cos^2\beta }{8 \pi
G_F m^2_{H^{\pm}}}\left( \frac{m_t }{v_u} -\epsilon ^{
u\,\star}_{33}\tan\beta \right)}
 { \left(\frac{m_t}{v_u} -\epsilon^u_{33}\tan\beta\right)  I_1(x),\nonumber\\
C_5^{(H^{\pm})}&=& - \frac { \sqrt{2}\,\alpha_s\cos^2\beta }{24
\pi G_F m^2_{H^{\pm}}} \left( \frac{m_t }{v_u} -\epsilon ^{
u\,\star}_{33}\tan\beta \right)}
 { \left(\frac{m_t}{v_u} -\epsilon^u_{33}\tan\beta\right)  I_1(x),\nonumber\\
C_6^{(H^{\pm})}&=&  \frac { \sqrt{2}\,\alpha_s\cos^2\beta }{8 \pi
G_F m^2_{H^{\pm}}} \left( \frac{m_t }{v_u} -\epsilon ^{
u\,\star}_{33}\tan\beta \right)}{ \left(\frac{m_t}{v_u} -\epsilon^u_{33}\tan\beta \right) I_1(x),\nonumber\\
C_7^{(H^{\pm})}&=&  \frac { \sqrt{2}\,\alpha\cos^2\beta }{6 \pi
G_F m^2_{H^{\pm}}} \left( \frac{m_t }{v_u} -\epsilon ^{
u\,\star}_{33}\tan\beta \right)} { \left(\frac{m_t}{v_u}
-\epsilon^u_{33}\tan\beta \right)\left( I_2(x)+I_3(x)\right),\nonumber\\
C_8^{(H^{\pm})}&=& 0,\nonumber\\
C_9^{(H^{\pm})}&=& \frac {\sqrt{2}\, \alpha\cos^2\beta }{6 \pi G_F
m^2_{H^{\pm}}} \left( \frac{m_t }{v_u} -\epsilon
^{u\,\star}_{33}\tan\beta \right)} { \left(\frac{m_t}{v_u}
-\epsilon^u_{33}\tan\beta \right)\left(
I_2(x)+I_3(x)-\frac{1}{\sin^2\theta_w}
 I_2(x)\right),\nonumber\\
C_{10}^{(H^{\pm})}&=& 0,\label{WilsL}
\end{eqnarray}
 Where the the loop functions $I_{1,2,3}(x)$
 are given by
\bea I_1(x)&=&
\frac{x\left(7x^2-29x+16\right)}{36\left(x-1\right)^3}
+\frac{x\left(3x-2\right)}{6\left(x-1\right)^4} \log{x}\eea and
\cite{Grossman:1999av}
\bea I_2(x)&=&\frac{x}{2(x-1)}-\frac{x}{2(x-1)^2}\log{x}\nonumber\\
I_3(x)&=& \frac{x\left(47x^2-79x+38\right)}{108\left(x-1\right)^3}
+\frac{x\left(-3x^2+6x-4\right)}{18\left(x-1\right)^4} \log{x}
 \eea with $x=m_{t}^2/m_{H^{\pm}}^2$.  In  Eq.(\ref{WilsL}), we neglected the
small contributions to the Wilson coefficients from the terms that
are proportional to  $\epsilon^{ u}_{13}$ and $\epsilon^{ u}_{23}$
due to the strong constraints on these parameters from  $b\to
d\gamma$  and $b\to s \gamma$ respectively arising at the one
loop-level \cite{Crivellin:2013wna}.

 The  charged Higgs mediation can give rise to new set of
  Wilson coefficients corresponding to flipping the chirality in
  the effective Hamiltonian from left to right:
\begin{eqnarray}
\widetilde{C}_{1,2}^{(H^{\pm})}&=&0,\nonumber\\
\widetilde{C}_3^{(H^{\pm})}&=& - \frac {
\sqrt{2}\,\alpha_s\sin^2\beta }{24 \pi G_F
 m^2_{H^{\pm}}} \left( \frac{m_b }{v_d} -
\epsilon^{ d}_{33}\tan\beta \right)}{ \left( \frac{m_s}{v_d}
-\epsilon^{ d\,\star}_{22}\tan\beta \right)  I_1(x),\nonumber\\
\widetilde{C}_4^{(H^{\pm})}&=&  \frac {
\sqrt{2}\,\alpha_s\sin^2\beta }{8 \pi G_F
 m^2_{H^{\pm}}} \left( \frac{m_b }{v_d} -
\epsilon^{ d}_{33}\tan\beta \right)}{ \left( \frac{m_s}{v_d}
-\epsilon^{ d\,\star}_{22}\tan\beta \right)  I_1(x),\nonumber\\
\widetilde{C}_5^{(H^{\pm})}&=& - \frac {
\sqrt{2}\,\alpha_s\sin^2\beta }{24 \pi G_F
 m^2_{H^{\pm}}} \left( \frac{m_b }{v_d} -
\epsilon^{ d}_{33}\tan\beta \right)}{ \left( \frac{m_s}{v_d}
-\epsilon^{ d\,\star}_{22}\tan\beta \right)  I_1(x),\nonumber\\
\widetilde{C}_6^{(H^{\pm})}&=&  \frac {
\sqrt{2}\,\alpha_s\sin^2\beta }{8 \pi G_F
 m^2_{H^{\pm}}} \left( \frac{m_b }{v_d} -
\epsilon^{ d}_{33}\tan\beta \right)}{ \left( \frac{m_s}{v_d}
-\epsilon^{ d\,\star}_{22}\tan\beta \right)  I_1(x),\nonumber\\
\widetilde{C}_7^{(H^{\pm})}&=&  \frac {
\sqrt{2}\,\alpha\sin^2\beta }{6 \pi G_F
 m^2_{H^{\pm}}} \left( \frac{m_b }{v_d} -
\epsilon^{ d}_{33}\tan\beta \right)}{ \left( \frac{m_s}{v_d}
-\epsilon^{ d\,\star}_{22}\tan\beta \right) \left( I_2(x)+
I_3(x)\right),\nonumber\\
\widetilde{C}_8^{(H^{\pm})}&=& 0,\nonumber\\
\widetilde{C}_9^{(H^{\pm})}&=& \frac {\sqrt{2}\, \alpha\sin^2\beta
}{6 \pi G_F
 m^2_{H^{\pm}}} \left( \frac{m_b }{v_d} -
\epsilon^{ d}_{33}\tan\beta \right)}{ \left( \frac{m_s}{v_d}
-\epsilon^{ d\,\star}_{22}\tan\beta \right)\left( I_2(x)+
I_3(x)-\frac{1}{\sin^2\theta_w} I_2(x)\right),\nonumber\\
\widetilde{C}_{10}^{(H^{\pm})}&=& 0,\label{WilsonR}
\end{eqnarray} As before, in the above equation,  we neglected the small
contributions to the Wilson coefficients from the terms that are
proportional to $\epsilon^{ d\,\star}_{32}$ and $\epsilon^{
d\,\star}_{12}$ due to the strong constraints on these parameters
from tree-level contributions to FCNC
process\cite{Crivellin:2013wna}.

  The charged Higgs mediation at tree level
can lead to the following weak effective Hamiltonian
 \be {\mathcal
H}_{eff}= \frac{ G_F}{\sqrt{2}}V^*_{us}V_{ub} \sum^{14}_{i=11}
C^H_i(\mu) Q^H_i(\mu),\ee where $C^H_i$ are the Wilson
coefficients obtained by perturbative QCD running from
$M_{H^{\pm}}$ scale to the scale $\mu$ relevant for hadronic decay
and $Q^H_i$ are the relevant local operators at low energy scale
$\mu\simeq m_b$. The operators can be written as %
\bea
Q^H_{11} &=& (\bar{u} P_L b)(\bar{s} P_R u),\nonumber\\
Q^H_{12} &=& (\bar{u} P_R b)(\bar{s} P_L u),\nonumber\\
Q^H_{13} &=& (\bar{u} P_L b) (\bar{s} P_L u),\nonumber\\
Q^H_{14} &=& (\bar{u} P_R b)(\bar{s} P_R u),
 \eea

And  the corresponding Wilson coefficients $C^H_i$ are given as
\begin{eqnarray}
C^H_{11} &=& \frac {\sqrt{2} }{ G_F V^*_{us}V_{ub}
 m^2_H} \bigg(\sum\limits_{j = 1}^3
{\cos\beta\, V^{\star}_{j2} \left( \frac{m_u }{v_u} \delta_{j1}-
\epsilon^{ u}_{j1}\tan\beta \right)}\bigg)\bigg( \sum\limits_{k=
1}^3 {\cos\beta\,V_{k3}} \left( \frac{m_u}{v_u}
\delta_{k1}-\epsilon^{ u\,\star}_{k1}\tan\beta
\right)\bigg),\nonumber\\
C^H_{12} &=& \frac {\sqrt{2} }{ G_F V^*_{us}V_{ub}
 m^2_H} \bigg(\sum\limits_{j = 1}^3
{\sin\beta\,V_{1j}  \left( \frac{m_b }{v_d} \delta_{j3}-
\epsilon^{ d}_{j3}\tan\beta \right)}\bigg)\bigg( \sum\limits_{k=
1}^3 {\sin\beta\,V^{\star}_{1k}} \left( \frac{m_s}{v_d}
\delta_{k2}-\epsilon^{ d\star}_{k2}\tan\beta
\right)\bigg),\nonumber\\
 C^H_{13} &=& \frac {\sqrt{2} }{  G_F V^*_{us}V_{ub}
 m^2_H} \bigg(\sum\limits_{j = 1}^3 {\cos\beta\, V_{j3} \left(
\frac{m_u }{v_u} \delta_{j1}- \epsilon^{ u\star}_{j1}\tan\beta
\right)}\bigg)\bigg( \sum\limits_{k= 1}^3
{\sin\beta\,V^{\star}_{1k}} \left( \frac{m_s}{v_d}
\delta_{k2}-\epsilon^{ d\star}_{k2}\tan\beta
\right)\bigg),\nonumber\\
C^H_{14} &=& \frac {\sqrt{2} }{ G_F V^*_{us}V_{ub}
 m^2_H}\bigg( \sum\limits_{k= 1}^3
{\cos\beta\,V^{\star}_{k2}} \left( \frac{m_u}{v_u}
\delta_{k1}-\epsilon^{ u}_{k1}\tan\beta
\right)\bigg)\bigg(\sum\limits_{j = 1}^3 {\sin\beta\,V_{1j} \left(
\frac{m_b }{v_d} \delta_{j3}- \epsilon^{ d}_{j3}\tan\beta
\right)}\bigg), \nonumber \\
 \label{Higgsw}
\end{eqnarray}

\section{Numerical results and analysis}
In order to estimate the enhancements in the full Wilson
coefficients $ C_7 $ and $ C_9 $ due to the charged Higgs
contribution we define the ratios: $R^{H^{\pm}}_i =
{|C^{{H^{\pm}}}_i|}/{|C^{SM}_i|}$ and
$\tilde{R}^{H^{\pm}}_i={|\widetilde{C}^{{H^{\pm}}}_i|}/{|C^{SM}_i|}$
for $i=7,9$ where $C_i$ are the SM Wilson coefficients. These
ratios will give us an indication about the magnitudes of the
charged Higgs Wilson coefficients compared to the SM ones and thus
can give a hint of the expected enhancement or reduction in the
branching ratios of our decay channels. We also define the ratios
 ${\mathcal R}^{\,M}_{b_i}= \big(BR^{SM+H^{\pm}}_{i}(\bar{B}_s\to\phi
M)-BR^{SM}_{i}(\bar{B}_s\to\phi
M)\big)/BR^{SM}_{i}(\bar{B}_s\to\phi M)$ where $M=\pi , \rho$, $i=
1,2$ refers to solutions $1,2$ for the SCET parameter space for
which the corresponding amplitudes are given in
Eqs.(\ref{pi1},\ref{pi2}) and $ BR^{SM+H^{\pm}}(\bar{B}_s\to\phi
M)$ and $BR^{SM}(\bar{B}_s\to\phi M)$ are the branching ratios
obtained when we consider the total contributions including
charged Higgs and the SM contributions alone respectively. These
ratios will give us the size of the enhancement or reduction to
the branching ratios of our decay modes compared to the
contribution from the SM.

\subsection{Two Higgs doublet model type-II}

We start by considering two Higgs doublets models type II. In this
case the Wilson coefficients can be obtained from
Eqs.(\ref{WilsL},\ref{WilsonR}) by setting $\epsilon^{
u}_{33}=\epsilon^{ d}_{22}=\epsilon^{ d}_{33}=0$.

The requirement for the top and bottom Yukawa interaction to be
perturbative results in a constraint on $\tan \beta$ namely, $
0.4\lsim \tan\beta \lsim 91$ \cite{Aoki:2009ha}. LEP has performed
a direct search for a charged Higgs in 2HDM type-II   and they
have set a lower limit on the mass of the charged Higgs boson of
80 GeV at 95\% C.L., with the process  $e^+ e^- \to H^+ H^-$ upon
the assumption  $BR(H^+ \to \tau^+ \nu) + BR(H^+ \to c \bar s) +
BR(H^+ \to A W^+) =1$ \cite{Abbiendi:2013hk}. If $BR(H^+ \to
\tau^+ \nu) =1$ the bound on the mass of the charged Higgs  is 94
GeV \cite{Abbiendi:2013hk}. Recent results on $ B\to \tau \nu$
obtained by BELLE \cite{Hara:2010dk} and BABAR
\cite{Aubert:2009wt}  have strongly improved the indirect
constraints on the charged Higgs mass in type II 2HDM
\cite{Baak:2011ze}:
\begin{equation}
m_{H^+}> 240 GeV \ \  at \ \ 95 \% CL
\end{equation}
Other experimental bounds can be  applied on the $(\tan \beta,
m_{H^\pm}) $ plane such as the bounds from $B\to X_s \gamma$
\cite{BB2,Crivellin:2013wna}, $B_s \to \mu^+\mu^-$, $B\to \tau
\nu$, $K\to \mu \nu/\pi\to \mu\nu$ \cite{Crivellin:2013wna} and
the bounds from ATLAS~\cite{ATLASICHEP} and CMS~\cite{CMSICHEP}
collaborations coming from $pp \to t \bar{t} \to b \bar{b} W^\mp
H^\pm (\to \tau \nu)$.

 We note from Eq.(\ref{WilsL}), after setting $\epsilon^{
u}_{33}=0$, that the dependency of the Wilson coefficients
${C}_{7,9}^{(H^{\pm})}$ are on $cos^2\beta/v^2_u=1/(v\,
tan\beta)^2$. Thus  small values of $\tan \beta$ these Wilson
coefficients ${C}_{7,9}^{(H^{\pm})}$ will blow up and can enhance
sizeably the branching ratios of the decay channels under
consideration. On the other hand we note from Eq.(\ref{WilsonR}),
after setting $\epsilon^{ d}_{22}=\epsilon^{ d}_{33}=0$, the
situation is reversed for $\widetilde{C}_{7,9}^{(H^{\pm})}$ as the
dependency in this case is on $\cot^2 \beta$ and thus large values
of $\tan \beta$  can enhance the branching ratios. In both cases
small values of charged Higgs mass are required.

\begin{figure}[tbhp]
\includegraphics[width=7cm,height=7cm]{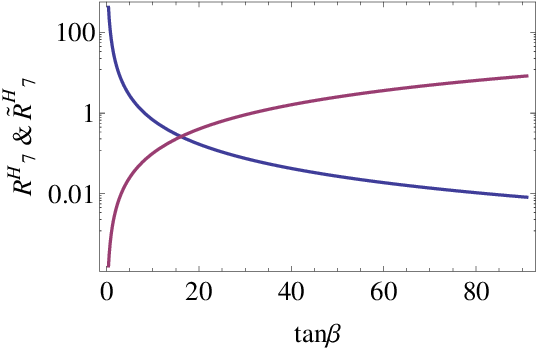}
\hspace{1.cm}
\includegraphics*[width=7cm,height=7cm]{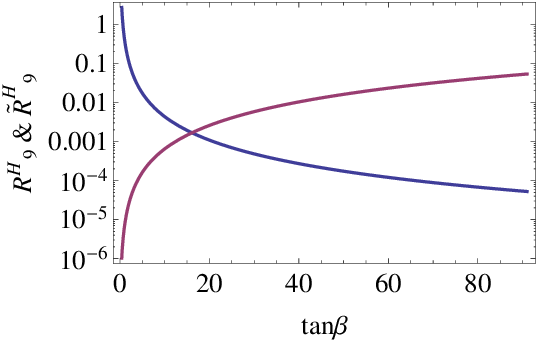}
\medskip
\caption{Left diagram corresponds to $R^{H^{\pm}}_7$
($\tilde{R}^{H^{\pm}}_7$) in units of $10^{-2}$ blue (red) curve
as a function of $\tan\beta$.  The right diagram corresponds to
$R^{H^{\pm}}_9$ ($\tilde{R}^{H^{\pm}}_9$) in units of $10^{-2}$
blue(red) curve as a function of $\tan\beta$. In both plots we
take $m_{H^{\pm}}=380\, GeV$.} \label{singlemas1}
\end{figure}

 In Fig.(\ref{singlemas1}) we plot
$R^{H^{\pm}}_i$ and $\tilde{R}^{H^{\pm}}_i$ for $i=7,9$ verses
$\tan{\beta}$ for a value of the charged Higgs mass
$m_{H^{\pm}}=380\, GeV$. This mass  is the lower limit of the
charged Higgs mass allowed by $B\to X_s \gamma$
constraints\cite{BB2}.  In the left diagram the blue (red) curve
corresponds to $R^{H^{\pm}}_7 (\tilde{R}^{H^{\pm}}_7)$ while in
the right diagram it corresponds to  $R^{H^{\pm}}_9
(\tilde{R}^{H^{\pm}}_9)$. As expected from Eq.(\ref{WilsL}) the
Wilson coefficients ${C}_{7,9}^{(H^{\pm})}$ vary  inversely with
$\tan^2 \beta$ which can is clear in Fig.(\ref{singlemas1}). Thus
larger values of ${C}_{7,9}^{(H^{\pm})}$ can be obtained for
smaller values of $\tan\beta$.  For a value of $\tan \beta = 0.4 $
allowed by the perturbativity of the top and bottom Yukawa
interaction we find that $R^{H^{\pm}}_7\simeq 400\%$.  This
indicates that ${C}_{7}^{(H^{\pm})}\simeq  4\, C^{SM}_7$ and
represent the maximum value can be reached as  $\tan \beta < 0.4 $
is excluded by the perturbativity of the  top and bottom Yukawa
interaction constraints. For the case of the Wilson coefficients
${C}_{9}^{(H^{\pm})}$ we find that $R^{H^{\pm}}_9\simeq 3\%$. This
indicates that ${C}_{9}^{(H^{\pm})}\simeq  0.03\, C^{SM}_9$.  For
larger values of $\tan\beta$ the ratios  $R^{H^{\pm}}_{7,9}$
become so small and close to zero as shown in
Fig.(\ref{singlemas1}) indicating very small values of the Wilson
coefficients ${C}_{7,9}^{(H^{\pm})}$ compared to their
corresponding ones in the SM . Turning now to the Wilson
coefficients $\widetilde{C}_{7,9}^{(H^{\pm})}$ where the
dependency in this case will be directly on $\tan^2 \beta$ as
shown in Eqs.(\ref{WilsonR}). Thus larger values of
$\widetilde{C}_{7,9}^{(H^{\pm})}$ can be obtained for larger
values of $\tan\beta$. They are represented by the red curves in
Fig.(\ref{singlemas1}).  For a value of $\tan \beta = 91 $ allowed
by the perturbativity of the top and bottom Yukawa interaction we
find that $\tilde{R}^{H^{\pm}}_7 \simeq 8\%$.  This indicates that
$\widetilde{C}_{7}^{(H^{\pm})}\simeq 0.08 \, C^{SM}_7$ and
represent the maximum value can be reached as  $\tan \beta > 91 $
is excluded by the perturbativity of the top and bottom Yukawa
interaction constraints. For the case of the Wilson coefficients
$\widetilde{C}_{9}^{(H^{\pm})}$ we find that $
\tilde{R}^{H^{\pm}}_9 \simeq 0.05\%$.  For smaller values of
$\tan\beta$ the ratios $\tilde{R}^{H^{\pm}}_{7,9}$ become so small
as shown in Fig.(\ref{singlemas1}) indicating very small values of
the Wilson coefficients $\widetilde{C}_{7,9}^{(H^{\pm})}$. We note
also from Fig.(\ref{singlemas1}) that $R^{H^{\pm}}_7 >>
R^{H^{\pm}}_9 $ and similarly for $\tilde{R}^{H^{\pm}}_7 >>
\tilde{R}^{H^{\pm}}_9$ this is because in the denominators of
these ratios  $C^{SM}_9 >> C^{SM}_7$.

\begin{figure}[tbhp]
\includegraphics[width=7cm,height=7cm]{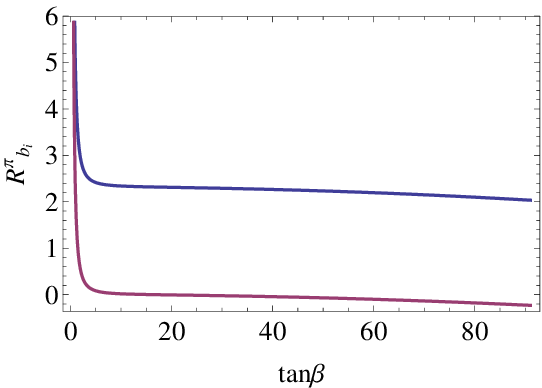}
\hspace{1.cm}
\includegraphics*[width=7cm,height=7cm]{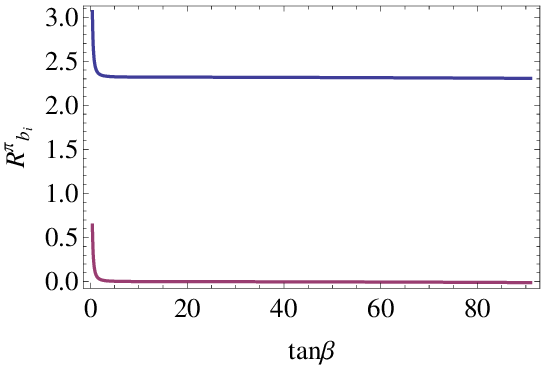}
\medskip
\caption{ ${\mathcal R}^{\pi}_{b_1}$ (${\mathcal R}^{\pi}_{b_2}$)
in units of $10^{-2}$ blue (red) curve as a function of
$\tan\beta$ for $m_{H^{\pm}}=380\, GeV$ left plot and the right
plot is for $m_{H^{\pm}}=1000\, GeV$.} \label{singlemas11}
\end{figure}

  Turning now to  the Wilson coefficients $C^H_{11}-C^H_{14}$
 given in Eq.(\ref{Higgsw}). By setting $\epsilon^{
u,d}_{ij}=0$ we find that $C^H_{11}$, $C^H_{13}$ and $C^H_{14}$
will be suppressed by the smallness of the product of quark masses
$m^2_u$, $m_u m_s$ and $m_u m_b$ respectively. For $C^H_{12}$ we
find that it is proportional to $m_s m_b\tan\beta$ which can be
enhanced for large values of $\tan\beta$ in a similar manner to
$C^D_{11}$ resulted from the charged Higgs mediation in the MSSM
with large $\tan\beta$ considered in Ref.\cite{Beneke:2009eb}.
Since all these Wilson coefficients have to be multiplied by the
CKM factor $\lambda^s_u$ they should be compared to the tree level
Wilson coefficient of the SM. Clearly $C^H_{11}$, $C^H_{13}$ and
$C^H_{14}$ can be safely drop and only $C^H_{12}$ can be
comparable with the SM tree level Wilson coefficients only when
$\tan\beta$ is large. However due to the constraints from $B^+\to
\tau^+\nu_{\tau}$, one find that  $C^H_{12}$ is roughly one order
of magnitude smaller than $C^{SM}_2$ as can be read from Eq.(24)
in Ref.\cite{Beneke:2009eb}. Thus we can also safely drop
$C^H_{12}$ in our analysis.

   In Fig.(\ref{singlemas11}) we plot ${\mathcal R}^{\pi}_{b_1}$
(${\mathcal R}^{\pi}_{b_2}$) , blue(red) curve,  as a function of
$\tan\beta$ for $m_{H^{\pm}}=380\, GeV$ and $ m_{H^{\pm}}=1000\,
GeV$. For the lower bound on $\tan\beta=0.4$ and for
$m_{H^{\pm}}=380\, GeV$ we find that ${\mathcal R}^{\pi}_{b_1}
\simeq 18\%$, ${\mathcal R}^{\pi}_{b_2} \simeq 14\%$ which means
charged Higgs contributions to the branching ratio of
$\bar{B}_s\to \phi\pi$ can reach a maximum value $18\%$ of the SM
prediction. For $m_{H^{\pm}}=1000\, GeV$ and $\tan\beta=0.4$ the
charged Higgs contributions to the branching ratio of
$\bar{B}_s\to \phi\pi$ can reach  $3\%$ and $0.64\%$ of the SM
prediction corresponding to solutions $1$ and $2$ of the SCET
parameter space respectively as shown in the plot. We note from
Fig.(\ref{singlemas11}) that ${\mathcal R}^{\pi}_{b_1} > {\mathcal
R}^{\pi}_{b_2}$ for all values of $\tan\beta$. This can be
explained by noticing that their denominators are
$BR^{SM}_1(\bar{B}_s\to\phi \pi)$ and $BR^{SM}_2(\bar{B}_s\to\phi
\pi)$ and  form Table \ref{branch} we have
$BR^{SM}_2(\bar{B}_s\to\phi \pi)> BR^{SM}_1(\bar{B}_s\to\phi \pi)
$. Another remark is that ${\mathcal R}^{\pi}_{b_2}$ varies with
$\tan\beta$ and can have positive, zero and negative values. The
reason is as follows: for $\tan\beta < 5$ we see from
Fig.(\ref{singlemas1}) that ${C}_{7}^{(H^{\pm})} \gg
\widetilde{C}_{7}^{(H^{\pm})}$. Note also ${C}_{7}^{(H^{\pm})}$
has similar sign to $C^{SM}_7$ and thus it leads to instructive
effect and enhance the amplitude.  For values of $ 5 < \tan\beta <
20$ we find that the term in the amplitude proportional to
$\widetilde{C}_{7}^{(H^{\pm})}$ starts to be non zero and have
opposite sign to the total Wilson coefficient $C_7$ leading to a
destructive effect and  almost Higgs contributions become
negligible and thus we get  ${\mathcal R}^{\pi}_{b_2}=0$. For $
\tan\beta\geq 20$ we find that $\widetilde{C}_{7}^{(H^{\pm})}>
{C}_{7}^{(H^{\pm})}$ and thus it reduces the amplitude leading to
$ BR^{SM+H^{\pm}}(\bar{B}_s\to\phi M) < BR^{SM}(\bar{B}_s\to\phi
M)$ and thus we obtain the negative values in the plot. Turning to
 ${\mathcal R}^{\pi}_{b_1}$ we find the effect caused by the relative size of
$\widetilde{C}_{7}^{(H^{\pm})}$ and ${C}_{7}^{(H^{\pm})}$ is small
as the coefficient of the $\widetilde{C}_{7}^{(H^{\pm})}$ term in
the amplitude corresponding to solution $1$  is  smaller than  its
corresponding one in solution $2$. This explains why we do not
have zero and negative values for  ${\mathcal R}^{\pi}_{b_1}$ as
we have for  ${\mathcal R}^{\pi}_{b_2}$ as shown in
Fig.(\ref{singlemas11}).

\begin{figure}[tbhp]
\includegraphics*[width=7.5cm,height=8.5cm]{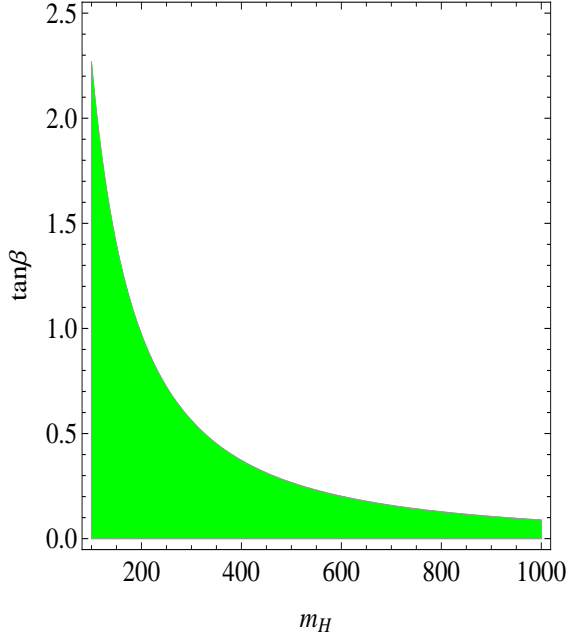}
\medskip
\caption{Allowed values of the parameter space which enhance  Br
$(\bar{B}_s\to \phi \pi^0) $ by more than or equal $18\%$ for
solution 1 of the SCET parameter space.} \label{singlemas3}
\end{figure}

So far we have applied only the constraints from the requirement
that the top and bottom Yukawa interaction to be perturbative to
just give an estimation of the maximum enhancement can be obtained
in 2HDMs type-II. We have selected two values of the charged Higgs
mass and found that for the two values of the charged Higgs mass
$m_{H^{\pm}}=380\,GeV$ and $m_{H^{\pm}}= 1000\, GeV$ the maximum
enhancement can be $18\%$ of the SM prediction and correspond to
solution $1$ of the SCET parameter space. Thus for charged Higss
masses smaller than $380\,GeV$ and vales of $ \tan\beta \lsim 0.4
$ the enhancement in the branching ratio of $\bar{B}_s\to \phi\pi$
can exceed $ 18\%$. This result motivates us to determine the
regions in the $(\tan \beta, m_{H^\pm}) $ plane which the
enhancement in the branching ratio of $\bar{B}_s\to \phi\pi$ can
be $18\%$ or more of the SM prediction. In Fig.(\ref{singlemas3})
we plot this region in the $(\tan \beta, m_{H^\pm}) $ plane.

 In Ref.\cite{Crivellin:2013wna}, see Figure 1, an updated
study of the possible constraints imposed on the $(\tan \beta,
m_{H^\pm}) $ plane of the two Higgs doublet model type-II from the
experimental measurements in $B\to s \gamma$, $B\to D\tau \nu$,
$B\to \tau \nu$, $K\to \mu \nu/\pi\to \mu\nu$, $B_s \to
\mu^+\mu^-$ and $B\to D^*\tau \nu$ showed that no region in the
$(\tan \beta, m_{H^\pm}) $ plane is compatible with all these
processes.  Explaining $B\to D^*\tau \nu$ requires large values of
$\tan\beta$ and very small Higgs mass and thus together with $B\to
s \gamma$ constraints excludes the green region in
Fig.(\ref{singlemas3}). Thus we conclude that the enhancement in
the branching ratio is always less than $18\%$ for the allowed
regions in the $(\tan \beta, m_{H^\pm})$ plane assuming no
constraints from the anomaly in $B\to D^*\tau \nu$ observed by
BABAR. However if this anomaly is confirmed in the near future by
other experiments, such as Belle-II  experiment, then taking into
account  $B\to D^*\tau \nu$ and $B\to s \gamma$ will rule out the
whole parameter space of the charged Higgs in the two Higgs
doublet model type-II.

  As can be seen from Table \ref{branch} the errors of the SM
predictions to the branching ratios are approximately $40\%$ and
thus it is clear that  the enhancement in the branching ratio by
$18\%$ with respect to the SM predictions due to the charged Higgs
mediation will be invisible within the theoretical uncertainties
in the SM predictions.

 Turning now to the branching ratios of $\bar{B}_s\to \phi\rho$, we
find that they  can be enhanced or reduced by the charged Higgs
contribution. However the enhancement or the reduction  are always
less than $4\%$  of the SM prediction for the allowed regions in
the $(\tan \beta, m_{H^\pm})$ plane.

\subsection{Two Higgs doublet model type-III}

We turn now to the case of two Higgs doublet models type III. In
this case the Wilson coefficients are those given in
Eqs.(\ref{WilsL},\ref{WilsonR}) and  the parameter space contains
extra parameters which are the couplings $\epsilon^{ q}_{ij}$
where $q=u,d$ appears in  the Yukawa Lagrangian.

 We start our analysis by  discussing the  constraints on the
parameters $ \epsilon^{u}_{33}$, $ \epsilon^{d}_{22}$ and $
\epsilon^{d}_{33}$ relevant to our decay modes.  Possible
constraints on these parameters can be obtained by applying the
naturalness criterion of 't Hooft to the quark masses. According
to this criterion the smallness of a quantity is only natural if a
symmetry is gained in the limit in which this quantity is zero
\cite{Crivellin:2012ye}. Hence applying this  criterion to the
quark masses in the 2HDM of type III we find that for $i\geq j$
\cite{Crivellin:2013wna}
\begin{eqnarray}
|v_{u(d)} \epsilon^{d(u)}_{ij}|\leq \,{\rm max
}\left[m_{d_i(u_i)},m_{d_j(u_j)}\right]\,.
\label{thooft}\end{eqnarray}

As can be seen from the above equation that $\epsilon^d_{22}$ will
be severely constrained by the small mass of the strange quark. In
addition the constraints are expected to become more stronger with
increasing the value of $\tan \beta$ due to the inverse dependency
on $v_u= v \sin\beta$ which increase with increasing $\tan \beta$.
However we find that $v_u$ changes slightly with varying
$\tan\beta$ and thus the constraints are insensitive to the values
of $\tan\beta$. It is easy to check that the absolute values of
the $\epsilon^{ d\,\star}_{22}\tan\beta $ are always very small in
comparison with the term ${m_s}/{v_d} $ for all values of
$\tan\beta$ and thus we can safely drop $\epsilon^{
d\,\star}_{22}\tan\beta $ terms  in Eq.(\ref{WilsonR}) comparing
to ${m_s}/{v_d} $.

\begin{figure}
  \includegraphics[width=7.5cm]{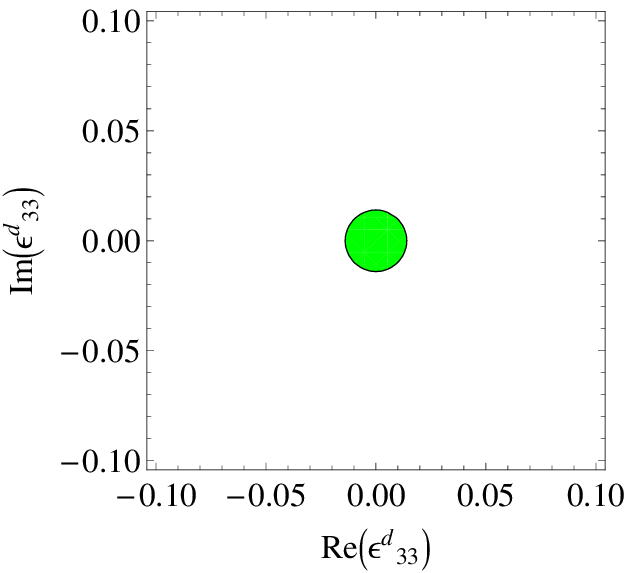}
  \includegraphics[width=7.5cm]{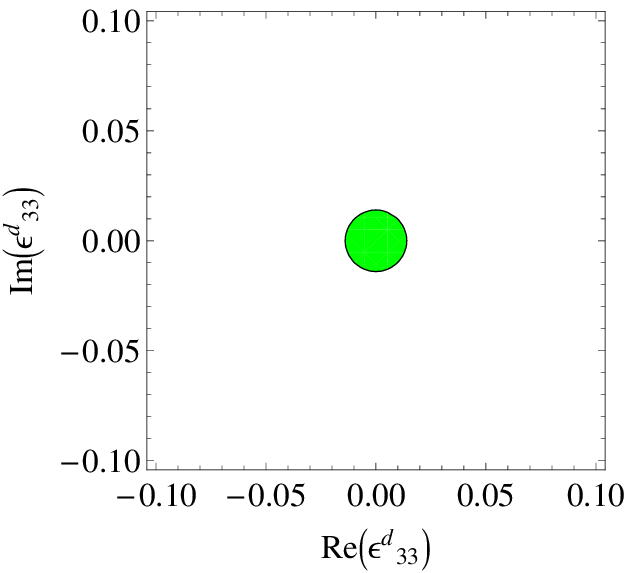}
  \caption{Constraints on  $\epsilon^d_{33}$ obtained upon applying the
naturalness criterion of 't Hooft to the quark masses. Left plot
corresponding to $\tan\beta =10$  while right plot corresponding
to $\tan\beta =50$.}\label{higsplane}
\end{figure}

 Turning now to $\epsilon^d_{33}$ we find also from
Eq.(\ref{thooft}) that it  is less constrained compared to
$\epsilon^d_{22}$ as the bottom quark mass is very large compared
to the mass of the strange quark. Moreover, we find that the
absolute values of  $\epsilon^{ d\,\star}_{22}\tan\beta $  are
still comparable with the term ${m_b}/{v_d} $ and thus we can not
drop these terms as we did for the case of $\epsilon^{
d\,\star}_{22}\tan\beta $ terms. In Fig.(\ref{higsplane}) we show
the allowed values of the real and imaginary parts of
$\epsilon^d_{33}$ corresponding to two different values of $\tan
\beta$. As can be seen from the figure that the constraints are
insensitive to varying $\tan\beta $ as we discussed above.

  The constraints imposed on $\epsilon^u_{33}$  by applying the
naturalness criterion of 't Hooft to the top quark mass is
expected to be even weaker than those obtained for
$\epsilon^d_{33}$ due to the so large top quark mass compared to
the bottom quark mass.  Moreover we expect that the constrains
becomes more loose with increasing the value of $\tan \beta$ due
to the inverse dependency on $v_d= v \cos\beta$ which decrease
significantly  with increasing $\tan \beta$.  Thus we can not rely
on the naturalness criterion of 't Hooft to constrain
$\epsilon^u_{33}$. In Ref.\cite{Crivellin:2013wna} an extensive
study of the flavor physics in the context of two Higgs doublet
model type-III has been performed to constrain the model both from
tree-level processes and from loop observables. It is shown that
possible  constraints on $\epsilon^u_{33}$ can be obtained from
$B_s-\bar{B}_s$ mixing and $B\rightarrow X_s\gamma$. Moreover the
constraints on $\epsilon^{ u}_{33}$ from $B\rightarrow X_s\gamma$
are the most important ones. For instance, applying $B\rightarrow
X_s\gamma$ constraints, for $ m_{H^\pm}=500\,GeV$ and $\tan
\beta=50$ the coupling $\epsilon^{ u}_{33}$ should satisfy $\mid
\epsilon^{ u}_{33}\mid\leq 0.55 $ and the constrains become more
strong for smaller values of $ m_{H^\pm}$ and large values of
$\tan \beta$. Thus in our analysis we take into account the
constraints imposed on $\epsilon^d_{33}$ and $\epsilon^u_{33}$
discussed in Ref.\cite{Crivellin:2013wna}.

 In 2HDMs type-III the constraints on the charged Higgs
 mass from  $B\to X_s \gamma$ become weaker comparing with their
 corresponding  constraints in 2HDMs type-II. This because the off-diagonal
 parameter  $\epsilon^u_{23}$ can lead to a destructive interference
 with the SM (depending on its phase) and thus reduces 2HDMs type-III
contribution to the amplitude \cite{Crivellin:2013wna}. Thus the
lower limit on the charged Higgs mass of $380\,GeV$ in 2HDMs
type-II can be pushed down in  2HDMs type-III.

We start by discussing the effects of the presence of the
$\epsilon^d_{33}$ terms on the Wilson coefficients
$\widetilde{C}_{7,9}^{(H^{\pm})}$. Since $\epsilon^d_{33}$ is
generally complex, we expect that these terms can enhance or
reduce $\widetilde{C}_{7,9}^{(H^{\pm})}$ comparing to their values
in the two Higgs doublet model type-II. For $\tan \beta=50$ and
$m_{H^\pm}=300\,GeV$ we find that $\tilde{R}^{H^{\pm}}_7$ varies
in the range $3\%-7\%$ for the allowed values of $\epsilon^d_{33}$
by the naturalness criterion of 't Hooft constraints. Setting the
real and imaginary parts of $\epsilon^d_{33}$ to zeros leads to
$\tilde{R}^{H^{\pm}}_7= 5\%$ which we would obtain in two Higgs
doublet model type-II. Thus the presence of $\epsilon^d_{33}$
terms would enhance or reduce $\widetilde{C}_{7}^{(H^{\pm})}$ by
$2\%$ only. For $\tan \beta=30$ and $m_{H^\pm}=300\,GeV$ we find
that the enhancement or reduction is almost $1\%$ while for $\tan
\beta=80$ the enhancement or reduction is almost $4\%$. For
$\tilde{R}^{H^{\pm}}_9$ we find that the enhancements or the
reductions are much smaller than the case of
$\tilde{R}^{H^{\pm}}_7$  since $C^{SM}_9>> C^{SM}_7$. As a result
we conclude that the enhancements or the reductions of the Wilson
coefficients $\widetilde{C}_{7,9}^{(H^{\pm})}$ due to the presence
of the  $\epsilon^d_{33}$ terms  are not significant compared to
the case of two Higgs doublet model type-II and they almost
neglige for values of $\tan \beta \leq 30$.

  We turn now to  discuss the effects of the presence of the
$\epsilon^u_{33}$ terms on the Wilson coefficients
${C}_{7,9}^{(H^{\pm})}$ in a similar way as we did for
$\epsilon^d_{33}$. Again as $\epsilon^u_{33}$ is generally complex
we expect that these terms can enhance or reduce
${C}_{7,9}^{(H^{\pm})}$ comparing with their values in the two
Higgs doublet model type-II. However since the allowed values for
$\epsilon^u_{33}$ by $B\rightarrow X_s\gamma$ constraints exclude
negative values of the real part of $\epsilon^u_{33}$, see figures
17 and 18 in Ref.\cite{Crivellin:2013wna}, we find that the
$\epsilon^u_{33}$ terms always enhance ${C}_{7,9}^{(H^{\pm})}$
comparing with their values within two Higgs doublet model
type-II. As before we expect the enhancements to be larger for the
Wilson coefficient ${C}_{7}^{(H^{\pm})}$ and thus we only focus on
${R}^{H^{\pm}}_7$ in the following discussion. For $\tan \beta=50$
and $m_{H^\pm}=300\,GeV$ we find that ${R}^{H^{\pm}}_7$ can reach
$ 13\%$  which means that ${C}_{7}^{(H^{\pm})}$ can reach $13\%$
of ${C}^{SM}_{7}$. Setting $\epsilon^u_{33}=0$ we obtain the value
${R}^{H^{\pm}}_7 < 1\%$ which is the limit within two Higgs
doublet model type-II. This indicates that the presence of
$\epsilon^u_{33}$ terms can enhance the value of ${R}^{H^{\pm}}_7$
within two Higgs doublet model type-II by $13\%$. For $\tan
\beta=30$ the constraints become weaker than the case of $\tan
\beta=50$ and thus we expect to have larger enhancement. In this
case we find that ${R}^{H^{\pm}}_7$ can reach $40\%$ indicating
that within two Higgs doublet model type-III,
${C}_{7}^{(H^{\pm})}$ can reach $40\%$ of ${C}^{SM}_{7}$. Setting
$\epsilon^u_{33}=0$  we obtain the value ${R}^{H^{\pm}}_7 \simeq
0.2\%$ which is the limit within two Higgs doublet model type-II.
Clearly, the presence of $\epsilon^u_{33}$ terms  enhance  the
value of ${R}^{H^{\pm}}_7$ from $0.2\%$ in 2HDMs type-II to $40\%$
in 2HDMs doublet type-III.

  For  the Wilson coefficients $C^H_{11}-C^H_{14}$
 given in Eq.(\ref{Higgsw}) and keeping the  $\epsilon^{
u,d}_{ij}$ parameters  we still  find that they are still
suppressed either by the smallness of the quark masses or the
constraints applied on the $\epsilon^{ u,d}_{ij}$ parameters
\cite{Crivellin:2013wna} and thus we  drop their contributions in
our analysis.

  Turning now to the branching ratios of $\bar{B}_s\to \phi \pi^0 $
and $\bar{B}_s\to \phi \rho^0$ we note from
Eqs.(\ref{pi1},\ref{pi2}) that an enhancement in ${C}_{7}$ will
enhance the branching ratio of $\bar{B}_s\to \phi \pi^0 $ and
reduce at the same time $\bar{B}_s\to \phi \rho^0 $ due to the
opposite sign of the terms proportional to ${C}_{7}$. Since the
enhancement is large for the case of $\tan \beta=30$ we find that
 ${\mathcal R}^{\pi}_{b_i}$ can be enhanced by about $4\%$ of the SM
prediction for solution 1 while for solution 2 it is still very
small about $1\%$. Comparing the branching ratio of $\bar{B}_s\to
\phi \pi^0 $ corresponding to solution $1$ in 2HDMs type-III with
its value in 2HDMs type-II we find that  ${\mathcal
R}^{\pi}_{b_1}$ is enhanced by about a factor $2$. For smaller
values of $\tan\beta$ where the constraints on $\epsilon^u_{33}$
becomes more weaker we find that the predictions for the branching
ratios are close to their values for $\tan\beta=30$  as
$\epsilon^u_{33}$ is multiplied by $\tan\beta$ and thus
enhancement in $\epsilon^u_{33}$ will not be significant when it
is multiplied by small value of $\tan\beta$. Thus the branching
ratios in 2HDMs type-III are approximately equal their values in
2HDMs type-II.
 For the case of $\bar{B}_s\to \phi \rho^0 $ we find that
the reductions by the presence of $\epsilon^{u,d}_{33}$ terms are
almost neglige. Thus we conclude that although the presence of
$\epsilon^{u,d}_{33}$ terms enhance the branching ratio of
$\bar{B}_s\to \phi \pi^0 $ by about a factor $2$ of their values
in 2HDMs type--II  still the enhancement is not sizable compared
to the SM predictions  and will be also invisible within the
theoretical uncertainties in the SM predictions for the branching
ratios as for the case of 2HDMs type-II.
\section{Conclusion \label{sec:conclusion}}
In this work we have studied the decay modes $\bar{B}_s\to \phi
\pi^0$ and $\bar{B}_s\to \phi \rho^0$ within the frameworks of
two-Higgs doublet models type-II and typ-III. We adopt in our
study SCET as a framework for the calculation of the amplitudes.
Within the framework of two-Higgs doublet models type-II and
typ-III the  charged Higgs boson can mediate the $b\rightarrow s$
transition at quark level and thus generate the decay modes  $
\bar{B}_s\to \phi \pi^0$ and $ \bar{B}_s\to \phi \rho^0 $.  We
have derived the contributions of the charged Higgs mediation to
the weak  effective Hamiltonian governing the decay processes and
calculated the corresponding Wilson coefficients in both models.
In addition we have analyzed the effect of the charged Higgs
mediation on the Wilson coefficients of the electrowek penguins
and on the branching ratios of $\bar{B}_s\to \phi \pi^0$ and
$\bar{B}_s\to \phi \rho^0$ decays.

  Within two-Higgs doublet models type-II and type-III
  we find that the Wilson coefficients $C_7$ and  $ C_9$ can
be enhanced due to the contributions from the charged Higgs
mediation. As a consequence the branching ratios of $ \bar{B}_s\to
\phi \pi^0$ and $ \bar{B}_s\to \phi \rho^0 $ decays are enhanced
in turn. Moreover we have shown that the charged Higgs mediation
can lead also to new set of Wilson coefficients obtained from the
weak effective Hamiltonian by changing the chirality from left to
right. The presence of these new Wilson coefficients can also lead
to enhancement of the branching ratios of $\bar{B}_s\to \phi
\pi^0$ and $\bar{B}_s\to \phi \rho^0$ decays.

   We have shown that, within two-Higgs doublet models type-II,
the enhancement in the branching ratio of $\bar{B}_s\to \phi
\pi^0$ can not exceed $18\%$  with respect to the SM predictions
for a charged Higgs mass $380\,GeV$. For the branching ratio of
$\bar{B}_s\to \phi \rho^0$, we find that the charged Higgs
contribution in this case is small where the branching ratio of
$\bar{B}_s\to \phi \rho^0$ can be enhanced or reduced by about 4\%
with respect to the SM predictions.

  Turning to two-Higgs doublet models type-III we have shown
for a value of the charged Higgs mass $300\,GeV$ and $\tan
\beta=30$ although the enhancement in BR ($\bar{B}_s\to \phi
\pi^0$) can be about a factor $2$ of its value within 2HDMs
type-II however it is only $4\%$ enhancement with respect to the
SM predictions. For smaller values of $\tan \beta$ the predictions
for the branching ratios are close to their predictions in 2HDMs
type-II.   We show also that, since the errors of the SM
predictions to the branching ratios are approximately $40\%$ for
$\bar{B}_s\to \phi \pi^0$, the enhancement in the branching ratios
due to the charged Higgs mediation will be invisible within the
theoretical uncertainties in the SM predictions. Clearly, charged
Higgs contributions can not lead to a significant enhancement of
the branching ratios of $\bar{B}_s\to \phi \pi^0$ and
$\bar{B}_s\to \phi \rho^0$ decays by one order of magnitude over
their SM predictions making them possible for detection at LHC.

 \section*{Acknowledgements}
  This work is  supported by the research grant NTU-ERP-102R7701 and by the grants NSC 99-
2112-M-008- 003-MY3, NSC 100-2811-M-008-036 and NSC 101-
2811-M-008-022 of the National Science Council of Taiwan. We would
like to thank Masaya Kohda for useful discussions.

\end{document}